\documentclass[11pt]{article}
\usepackage{cite}
\usepackage{mdframed}
\usepackage{epsfig,subfigure}
\usepackage{amssymb}
\usepackage[fleqn]{amsmath}
\usepackage{color}
\usepackage{arydshln}
\def\QED{\mbox{\rule[0pt]{1.5ex}{1.5ex}}}

\usepackage{graphicx}
\usepackage{color}
\usepackage[dvipsnames]{xcolor}

\definecolor{armygreen}{rgb}{0.29, 0.33, 0.13}

\usepackage{amssymb}
\usepackage{amsmath,amsfonts,amssymb}
\usepackage{verbatim}
\usepackage{stfloats}
\usepackage[bookmarks=false]{}
\usepackage{algorithm}
\usepackage{algcompatible}

\newtheorem{theorem}{Theorem}

\newtheorem{lemma}{Lemma}

\newtheorem{remark}{Remark}

\newtheorem{example}{Example}

\newcommand\blfootnote[1]{%
  \begingroup
  \renewcommand\thefootnote{}\footnote{#1}%
  \addtocounter{footnote}{-1}%
  \endgroup
}

\begin{document}
\date{}

\title{
Compound Secure Groupcast: \\
Key Assignment for Selected Broadcasting
}
\author{\normalsize Hua Sun \\
}

\maketitle

\blfootnote{
Hua Sun (email: hua.sun@unt.edu) is with the Department of Electrical Engineering at the University of North Texas. }

\maketitle

\begin{abstract}
The compound secure groupcast problem is considered, where the key variables at $K$ receivers are designed so that a transmitter can securely groupcast a message to {\em any} $N$ out of the $K$ receivers through a noiseless broadcast channel. The metric is the information theoretic tradeoff between key storage $\alpha$, i.e., the number of bits of the key variable per message bit, and broadcast bandwidth $\beta$, i.e., the number of bits of the broadcast information per message bit.

We have three main results. First, when broadcast bandwidth is minimized, i.e., when $\beta = 1$, we show that the minimum key storage is $\alpha = N$. Second, when key storage is minimized, i.e., when $\alpha = 1$, we show that broadcast bandwidth $\beta = \min(N, K-N+1)$ is achievable and is optimal (minimum) if $N=2$ or $K-1$. Third, when $N=2$, the optimal key storage and broadcast bandwidth tradeoff is characterized as $\alpha+\beta \geq 3, \alpha \geq 1, \beta \geq 1$.
\end{abstract}

\newpage

\allowdisplaybreaks
\section{Introduction}
Secure groupcast \cite{Sun_SecureGroupcast} seeks the most efficient solution to communicate with a group of receivers over a noiseless broadcast channel securely such that the remaining receivers do not learn anything about the desired communication. The primary enabler for secure groupcast is that each receiver is equipped with a correlated key and the transmitter needs to exploit the keys available at the qualified receivers for group communication. At the same time, the keys available to the eavesdropping receivers are the fundamental challenge as each of the eavesdropping receivers also has a correlated key and we need to prevent leakage under such multiple intertwined views.

\begin{figure}[h]
\begin{center}
\includegraphics[width=7 in]{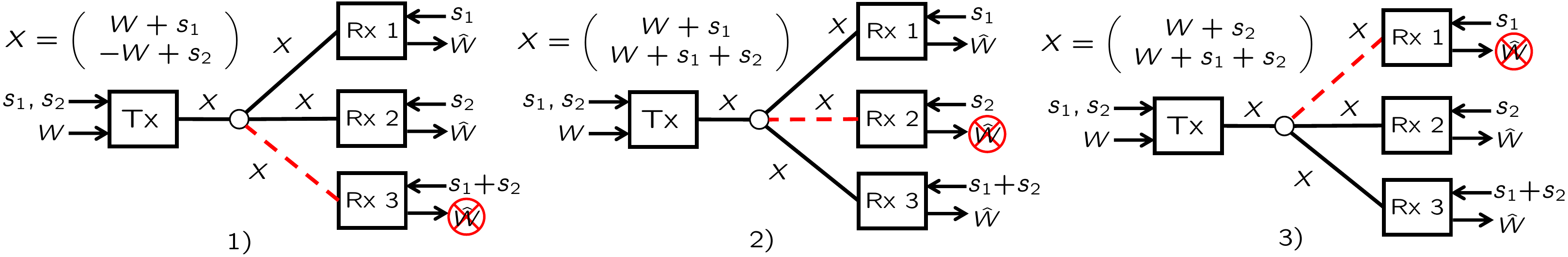}
\caption{\small Compound secure groupcast to any $N=2$ of $K=3$ receivers. $s_1, s_2$ are $2$ uniform i.i.d. symbols from any finite field. Keys are assigned as $Z_1 = s_1, Z_2 = s_2, Z_3 = s_1+s_2$. The secure groupcast schemes are shown when 1) the first $2$ receivers, 2) Receiver 1 and Receiver 3, and 3) the last $2$ receivers are qualified. Note that the key assignment remains the same for all $3$ scenarios.}
\label{fig:32}
\end{center}
\end{figure}

In the basic model of secure groupcast \cite{Sun_SecureGroupcast}, the key variables are fixed and given, e.g., the joint distribution and the sizes are not subject to design, and the identities of the external eavesdropping receivers are known globally. In this work, we relax the above two assumptions and consider the compound secure groupcast problem, where we may design the key variables $Z_k$ of the $K$ receivers (i.e., how to assign the keys is our choice) such that we can communicate a message $W$ with any $N$ ($1 \leq N \leq K-1$) out of the $K$ receivers in a secure manner by broadcasting $X$ (i.e., we do not know which receivers are qualified or eavesdropping beforehand), where the remaining $K-N$ receivers are ignorant of the desired message. An example of $N=2, K=3$ is shown in Fig.~\ref{fig:32}.

\vspace{0.1in}
\vspace{-0.1in}

The compound secure groupcast problem models an interesting and challenging scenario where the eavesdropping receivers are internal and their identities are not known in the key set-up stage. The applications can range from pay TV where the keys are distributed by a central controller and the message represents a particular channel subscribed by prime users, to a secure broadcasting system of an organization where the internal users may be compromised and are classified as unqualified users when sensitive information is later encrypted and heard by all users. 

To understand the fundamental limits of compound secure groupcast, we are interested in the following two metrics.
\begin{itemize}
\item Key Storage ($\alpha$) - To communicate $1$ bit of desired message $W$ (to any $N$ users), how many bits of the key $Z_k$ need to be stored at each receiver, denoted as $\alpha$? For example, consider Fig.~\ref{fig:32}, where $\alpha = 1$ as the key size is the same as the message size.
\item Broadcast Bandwidth ($\beta$) - To communicate $1$ bit of desired message $W$ (to any $N$ users), how many bits of the broadcast information $X$ need to be sent, denoted as $\beta$? For example, consider Fig.~\ref{fig:32}, where $\beta = 2$ as $2$ bits are broadcast to groupcast $1$ bit of desired message.
\end{itemize}

There exists an interesting tradeoff between key storage $\alpha$ and broadcast bandwidth $\beta$, i.e., if we are allowed to store more key symbols (when $\alpha$ is large), then we may send less information in the broadcast stage ($\beta$ can be small), and vice-versa, i.e., if the broadcast resource is abundant, then we may need to store fewer key symbols. The main motivation of this work is to understand such $(\alpha, \beta)$ tradeoff in the information theoretic sense.

Next, we summarize the main results obtained. We mainly focus on the extreme points where key storage is minimized, i.e., $\alpha = 1$ and where broadcast bandwidth is minimized, i.e., $\beta = 1$, and wish to characterize the minimum resource required for the other parameter. For both extreme points, there are natural feasible solutions. 

\vspace{0.05in}
\begin{figure}[h]
\begin{center}
\includegraphics[width=6 in]{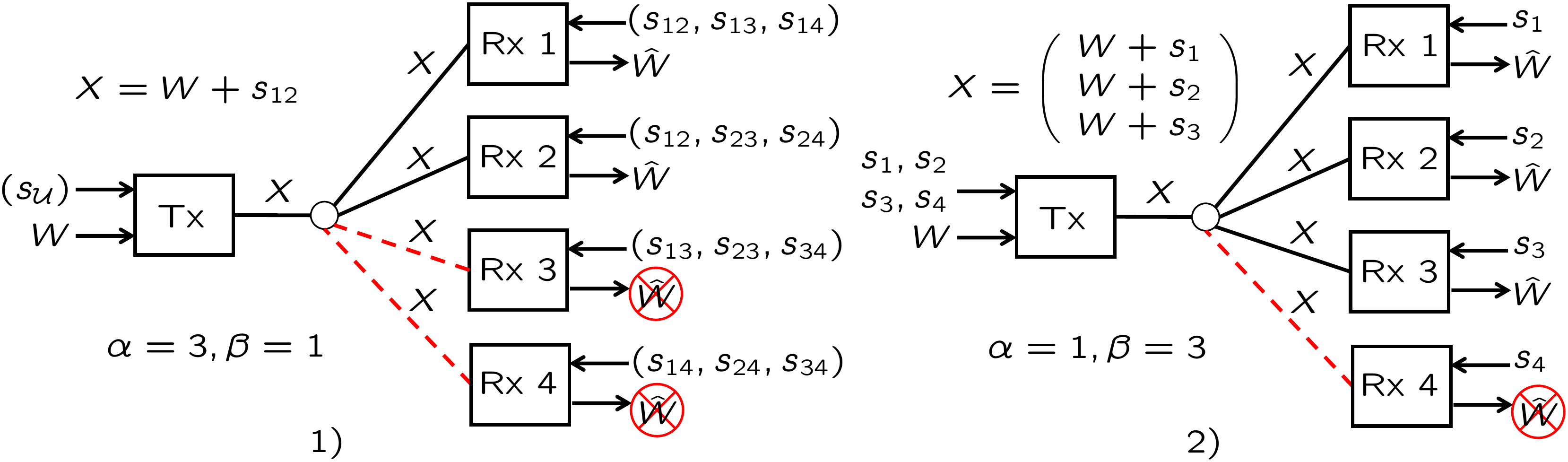}
\caption{\small The $s$ variables are independent. The key assignment is shown and transmitter knows all keys. 1) $N=2, K=4$. $\beta=1$ is minimized and a scheme of $\alpha=3$ is shown. 2) $N=3, K=4$. $\alpha=1$ is minimized and a scheme of $\beta=3$ is shown. One possible set of qualified receivers is plotted and other cases are similar.}
\label{fig:mot}
\end{center}
\end{figure}
\vspace{-0.3in}

\begin{itemize}
\item When $\beta = 1$, we share an independent key for any size $N$ subset of receivers so that for any set of qualified receivers, we use their exclusive shared key to send the desired message. The key storage achieved is $\alpha = \binom{K-1}{N-1}$. See Fig.~\ref{fig:mot}.1 for an example when $N=2, K=4$.
\item When $\alpha = 1$, we set the keys to be independent, i.e., $Z_k = s_k, \forall k \in \{1,\cdots,K\}$ and $s_k$ are uniform i.i.d. symbols. When any set of $N$ receivers are qualified, we simply send $W + s_q$ for every qualified Receiver $q$. The broadcast bandwidth achieved is $\beta = N$. See Fig.~\ref{fig:mot}.2 for an example when $N=3, K=4$.
\end{itemize}

Interestingly, we show that both schemes above are not optimal.
\begin{itemize}
\item When $\beta = 1$, we can achieve $\alpha = N$ with a generic vector linear scheme. Further, $\alpha = N$ is the optimal information theoretic key storage, i.e., the minimum achievable by any linear or non-linear schemes. For example, for the $N=2, K=4$ setting in Fig.~\ref{fig:mot}.1, $\alpha$ can be $2 < 3$ and the scheme is presented in Example \ref{ex:b1}. The general result is presented in Theorem \ref{thm:b1}.

\item When $\alpha = 1$, we can achieve $\beta = \min(N, K- N+1)$ with a generic scalar linear scheme. Further, $\beta = \min(N, K-N+1)$ is the optimal broadcast bandwidth when $N=2$ or $K-1$. For example, for the $N=3, K=4$ setting in Fig.~\ref{fig:mot}.2, $\beta$ can be $2 < 3$ and the scheme is presented in Example \ref{ex:a1}. The general result is presented in Theorem \ref{thm:a1}.
\end{itemize}

Finally, the insights obtained from above results allow us to characterize the optimal $(\alpha, \beta)$ tradeoff when $N=2$ and $K$ is arbitrary, as $\alpha + \beta \geq 3, \alpha \geq 1, \beta \geq 1$. Refer to Theorem \ref{thm:ab2}.

{\it Notation: For positive integers $K_1, K_2, K_1 \leq K_2$, we use the notation $[K_1:K_2] = \{K_1, K_1+1,\cdots, K_2\}$. The notation $|\mathcal{Q}|$ is used to denote the cardinality of a set $\mathcal{Q}$. 
The notation $\mathcal{A}\backslash\mathcal{B}$ denotes the difference of sets $\mathcal{A}, \mathcal{B}$, i.e., the set of elements that are in $\mathcal{A}$ but not in $\mathcal{B}$. In this work, a vector ${\bf v}$ denotes a row vector by default and ${\bf v}^T$ represents the transpose of ${\bf v}$, i.e., ${\bf v}^T$ is a column vector. For a matrix ${\bf V}$, the notation ${\bf V}(i, :)$ is used to denote the $i$-th row of ${\bf V}$ and the notation ${\bf V}(\mathcal{Q},:)$ is used to denote the sub-matrix of ${\bf V}$ formed by retaining only the rows with indices in the vector formed by arranging elements of the set of numbers $\mathcal{Q}$ in an increasing order.
}

\section{Problem Statement}\label{sec:model}
The compound secure groupcast problem has two stages - the key assignment stage and the secure groupcast stage, specified as follows.

In the key assignment stage, we design $K$ key variables $Z_1, \cdots, Z_K$, each of which consists of $L_Z$ symbols from a finite field\footnote{To simplify the presentation of the coding scheme, we allow a free choice of the field size $p$. This is consistent with an information theoretic formulation, where the actual size of the message is allowed to approach infinity and performance metrics are defined as ratios such that the effect of the field size is normalized (refer to (\ref{ab})).} $\mathbb{F}_p$ for a prime power $p$. $Z_k, k \in [1:K]$ is given to Receiver $k$.

In the secure groupcast stage, a transmitter must be able to send a message $W$ securely to any $N \in [1:K-1]$ receivers. The message $W$ consists of $L_W$ uniform i.i.d. symbols from $\mathbb{F}_p$ and is independent of the key variables (because e.g., $W$ is available after the keys are assigned).
\begin{eqnarray}
&& H(W) = L_W ~\mbox{(in $p$-ary units)}, \label{h1}\\
&& I(W; Z_1, \cdots, Z_K) = 0. \label{wz_ind}
\end{eqnarray}
The entropy function is measured in $p$-ary units throughout this paper.

When the message $W$ is securely groupcast to receivers in the set $\mathcal{Q} \subset [1:K], |\mathcal{Q}| = N$, the transmitter broadcasts signal $X_{\mathcal{Q}}$, where $X_{\mathcal{Q}}$ consists of $L_{X}$ symbols\footnote{We assume the scheme is symmetric, i.e., the size of the broadcast information $X_{\mathcal{Q}}$ does not depend on $\mathcal{Q}$, which has no loss of generality as any asymmetric scheme 
can be transformed to a symmetric one 
(through space sharing over permutations of the asymmetric scheme). 
Equivalently, we measure the broadcast information size using the worst case. The situation with the key size is similar, i.e., we assume the length of $Z_k$ does not depend on $k$.} from $\mathbb{F}_p$ and is heard perfectly by every Receiver $k$. For any Receiver $q$ that belongs to the qualified set $\mathcal{Q}$, the message must be recovered with no error\footnote{As the achievable schemes in this work all have zero error and zero leakage, for simplicity we do not adopt the vanishing error and leakage framework, under which our converse results do hold.},
\begin{eqnarray}
\mbox{[Correctness]}~~ H(W | X_{\mathcal{Q}}, Z_q) = 0, \forall q \in \mathcal{Q}. \label{corr}
\end{eqnarray}
For any Receiver $e$ that does not belong to the qualified set $\mathcal{Q}$, i.e., $e \in [1:K]\backslash\mathcal{Q}$, no information about the message shall be revealed.
\begin{eqnarray}
\mbox{[Security]}~~ I(W ; X_{\mathcal{Q}}, Z_e) = 0, \forall e \in [1:K]\backslash\mathcal{Q}. \label{sec}
\end{eqnarray}
We use the normalized key size (referred to as key storage and denoted by $\alpha$) and the normalized broadcast information size (referred to as broadcast bandwidth and denoted by $\beta$) to measure the performance, defined as follows.
\begin{eqnarray}
\alpha \triangleq \frac{L_Z}{L_W}, ~~ 
\beta \triangleq \frac{L_X}{L_W}. \label{ab}
\end{eqnarray}
A key storage and broadcast information tuple $(\alpha, \beta)$ is said to be achievable if there exists a compound secure groupcast scheme (i.e., a design of the key variables $Z_k$ and the broadcast signal variables $X_{\mathcal{Q}}$) such that the correctness constraint (\ref{corr}) and the security constraint (\ref{sec}) are satisfied for any $\mathcal{Q} \subset [1:K], |\mathcal{Q}| = N$, and key storage and broadcast bandwidth are smaller than or equal to $\alpha$ and $\beta$, respectively. The closure of the set of achievable $(\alpha, \beta)$ tuples is called the capacity region $\mathcal{C}$.

\section{Results}
In this section, we present our results along with illustrative examples and observations.

Let us start with a useful known converse result\footnote{The setting studied in \cite{Sun_SecureGroupcast} is secure groupcast, whose definitions are slightly different from the compound setting, e.g., broadcast bandwidth is defined as the scaling with respect to the key block length in secure groupcast, so we give a self-contained proof of Theorem \ref{thm:con} in Section \ref{sec:con} for completeness.}, borrowed from \cite{Sun_SecureGroupcast}.
\begin{theorem}\label{thm:con}
{\normalfont(Theorem 1 and Theorem 2 in \cite{Sun_SecureGroupcast})} For the compound secure groupcast problem (to any $N \in [1:K-1]$ of $K$ receivers), we have
\begin{eqnarray}
&& L_W \leq H(Z_q|Z_e), ~\forall q,e \in [1:K], q\neq e,  \label{rate_con} \\
&& L_X \geq |\mathcal{Q}| L_W - \left(\sum_{i=1}^{|\mathcal{Q}|}H(Z_{q_i}) - H(Z_{q_1}, \cdots, Z_{q_{|\mathcal{Q}|}}) \right), \notag\\
&&~~~\forall \mathcal{Q} =\{q_1, \cdots, q_{|\mathcal{Q}|}\} \subset [1:K], |\mathcal{Q}| \leq N. \label{com_con}
\end{eqnarray}
\end{theorem}

The extreme values of the key storage parameter $\alpha$ and the broadcast bandwidth parameter $\beta$ follow immediately from Theorem \ref{thm:con}.
\begin{eqnarray}
&& (\ref{rate_con}) \Rightarrow  L_W \leq H(Z_q) \leq L_Z \Rightarrow \alpha = L_Z/L_W \geq 1 \label{eq:e1}\\
&& (\ref{com_con}) \overset{|\mathcal{Q}|=1}{\Longrightarrow}  L_X \geq L_W \Rightarrow \beta = L_X/L_W \geq 1. \label{eq:e2} 
\end{eqnarray}
That is, the minimum key storage is $\alpha = 1$ and the minimum broadcast bandwidth is $\beta = 1$.
Next we proceed to our results on the corresponding extreme points of the $(\alpha, \beta)$ region.

\subsection{The Minimum Broadcast Bandwidth Extreme Point}
The optimal (minimum) key storage $\alpha$ for the minimum broadcast bandwidth extreme point ($\beta =1$) is characterized in the following theorem.
\begin{theorem}\label{thm:b1}
{\normalfont[Minimum $\alpha$ for Minimum $\beta$]}
For the compound secure groupcast problem (to any $N \in [1:K-1]$ of $K$ receivers), when broadcast bandwidth is minimized, $\beta = 1$, the 
minimum key storage is $\alpha = N$.
\begin{eqnarray}
\min\{ \alpha~ |~ (\alpha, \beta = 1) \in \mathcal{C}\} = N.
\end{eqnarray}
\end{theorem}

The proof of Theorem \ref{thm:b1} is presented in Section \ref{sec:b1} and we give an overview here. 

For the converse, it suffices to consider only $N+1 \leq K$ receivers, say receivers $1$ to $N+1$. Focus on any receiver, say Receiver $N+1$, who may be a part of $N$ different sets of qualified receivers, i.e., $\mathcal{Q}_1 = [1:N+1]\backslash\{1\}, \cdots, \mathcal{Q}_N = [1:N+1]\backslash\{N\}$. To securely groupcast the message to these $N$ sets of receivers, the transmit signals are $X_{\mathcal{Q}_1}, \cdots, X_{\mathcal{Q}_N}$. When $\beta = 1$, i.e., the transmit signal size is the same the message size, the essence is to realize that we can only use one-time pad and each of the transmit signal shall contain an independent key (formalized in Lemma \ref{lemma:b1}). After this claim is established in the information theoretic sense (more precisely, conditioned on the message $W$, the mutual information between $X_{\mathcal{Q}_1}, \cdots, X_{\mathcal{Q}_N}$ and $Z_{N+1}$ is no less than $N$ times the message size), we obtain that $\alpha \geq N$.

For the achievability, the key of each receiver consists of $N$ generic linear combinations of $N+1$ uniform i.i.d. basis key symbols, i.e., we operate over an $N+1$ dimensional key space and each receiver is assigned an $N$ dimensional subspace. As such, every $N$ receivers have an overlap of $N\times N - (N-1)(N+1) = 1$ dimension in generic position, which serves as the exclusive key when these $N$ receivers are qualified. As this overlap is determined only by the $N$ qualified receivers, this key (and the one-time pad transmit signal) is independent of the key known by any eavesdropping receiver (an $N$ dimensional generic subspace) such that the scheme is secure. Interestingly, elegant and efficient constructions over small field sizes exist for such generic spaces and an example is given below to illustrate the main idea.

\begin{example}\label{ex:b1}
Continuing from Fig.~\ref{fig:mot}.1, we consider the compound secure groupcast problem when $N=2, K=4$ and give a scheme that achieves $\alpha = 2, \beta = 1$. Following the intuitions presented above, we will assign a generic $2$ dimensional subspace over a $3$ dimensional space to each receiver when we set the keys. That is, define $s_1, s_2, s_3$ as $3$ uniform i.i.d. symbols from $\mathbb{F}_p$ (the value of $p$ will be specified later) and the key assignments are
\begin{eqnarray} 
Z_1 = {\bf V}_1  [s_1, s_2, s_3]^T, ~Z_2 = {\bf V}_2  [s_1, s_2, s_3]^T, ~Z_3 = {\bf V}_3  [s_1, s_2, s_3]^T, ~Z_4 = {\bf V}_4  [s_1, s_2, s_3]^T
\end{eqnarray}
where each of ${\bf V}_k, k \in \{1,2,3,4\}$ is a $2 \times 3$ matrix over $\mathbb{F}_p$ such that for any distinct elements $q_1, q_2, e_1, e_2$ from $\{1,2,3,4\}$,
\begin{eqnarray}
&& \dim(\mbox{rowspan}({\bf V}_{q_1}) \cap \mbox{rowspan}({\bf V}_{q_2})) = 1,\\
&& \mbox{rowspan}({\bf V}_{q_1}) \cap \mbox{rowspan}({\bf V}_{q_2}) ~\mbox{is independent of}~ \mbox{rowspan}({\bf V}_{e_1}),\\
&& \mbox{rowspan}({\bf V}_{q_1}) \cap \mbox{rowspan}({\bf V}_{q_2}) ~\mbox{is independent of}~ \mbox{rowspan}({\bf V}_{e_2}).
\end{eqnarray}
The above constraints are easily satisfied by random matrices, e.g., when every element of ${\bf V}_k$ is drawn independently and uniformly from $\mathbb{F}_p$ for a sufficiently large $p$ (proved by Schwartz-Zippel lemma \cite{Schwartz, Zippel, Demillo_Lipton}). However, an elegant solution exists due to Blom \cite{Blom1984} and is presented below (using a slightly different yet equivalent description that aligns with our intuition of generic spaces). Set $p = 5$ and consider $4$ distinct elements $v_1, v_2, v_3, v_4$ from $\mathbb{F}_p$, e.g., we may set $v_1 = 1, v_2 = 2, v_3 = 3, v_4 = 4$.
\begin{eqnarray}
{\bf V}_1 = \left[
\begin{array}{ccc}
1 & v_1 & 0 \\
0 & 1 & v_1
\end{array}
\right],
{\bf V}_2 = \left[
\begin{array}{ccc}
1 & v_2 & 0 \\
0 & 1 & v_2
\end{array}
\right],
{\bf V}_3 = \left[
\begin{array}{ccc}
1 & v_3 & 0 \\
0 & 1 & v_3
\end{array}
\right],
{\bf V}_4 = \left[
\begin{array}{ccc}
1 & v_4 & 0 \\
0 & 1 & v_4
\end{array}
\right].
\end{eqnarray}
The generic overlaps are as follows. For example, consider $\mbox{rowspan}({\bf V}_{1}) \cap \mbox{rowspan}({\bf V}_{2})$. The vector 
\begin{eqnarray}
{\bf v}_{\{1,2\}} = [1 ~~v_1+v_2 ~~v_1v_2]
\end{eqnarray}
is a linear combination of the rows of both ${\bf V}_1$ and ${\bf V}_2$, e.g., the first row plus $v_2$ times the second row for ${\bf V}_1$. Other choices of two ${\bf V}_k$ matrices are similar. Now when Receiver $1$ and Receiver $2$ are qualified, the transmit signal is
\begin{eqnarray}
X_{\{1,2\}} = W + {\bf v}_{\{1,2\}} [s_1, s_2, s_3]^T.
\end{eqnarray}
Correctness holds as ${\bf v}_{\{1,2\}}$ can be obtained locally at both Receiver $1$ and Receiver $2$. To guarantee security, we need that ${\bf v}_{\{1,2\}}$ is independent of ${\bf V}_3$ and ${\bf V}_4$, respectively. For example, consider ${\bf V}_3$.
\begin{eqnarray}
{\bf V}_{\{1,2\}\cup\{3\}} = \left[
\begin{array}{c}
{\bf V}_3 \\
{\bf v}_{12} 
\end{array}
\right] = \left[
\begin{array}{ccc}
1 & v_3 & 0\\
0 & 1 & v_3\\
1 & v_1 + v_2 & v_1v_2
\end{array} 
\right] \Rightarrow \det({\bf V}_{\{1,2\}\cup\{3\}}) = (v_3 - v_1) (v_3 - v_2) \neq 0 \label{eq:v123}
\end{eqnarray}
as $v_k$ are distinct.
\begin{eqnarray}
I(W; X_{\{1,2\}}, Z_3) &=& H(X_{\{1,2\}}, Z_3) - H(X_{\{1,2\}}, Z_3 | W) \\
&\overset{(\ref{wz_ind})}{=}& H(X_{\{1,2\}}, Z_3) - H({\bf V}_{\{1,2\}\cup\{3\}} [s_1, s_2, s_3]^T) \label{} \\
&\leq& 3 - 3 = 0
\end{eqnarray}
where in the last step, the first term follows from the fact that $X_{\{1,2\}}, Z_3$ consists of only $3$ symbols from $\mathbb{F}_p$ and uniform distributions maximize entropy, and the second term follows from (\ref{eq:v123}) stating that ${\bf V}_{\{1,2\}\cup\{3\}}$ has full rank. Hence, the security constraint (\ref{sec}) is satisfied.
Finally, the achieved performance is as desired because $\alpha = 2 = N$ ($2$ key symbols are stored for a $1$ symbol message), and $\beta = 1$ ($1$ symbol is broadcast to groupcast a $1$ symbol message).

Interestingly, the generic space assignment when $N=2$ introduced by Blom is generalized to arbitrary $N$ by Matsumoto and Imai \cite{Matsumoto_Imai}, in the context of key predistribution. The details can be found in Section \ref{sec:b1} and the connection of compound secure groupcast to prior work in key predistribution (and other problems) is discussed in Section \ref{sec:prior}.
\end{example}

\subsection{The Minimum Key Storage Extreme Point}
A scalar linear achievable scheme for the minimum key storage extreme point ($\alpha = 1$) is presented in the following theorem. 

\begin{theorem}\label{thm:a1}
{\normalfont[Achievable $\beta$ for Minimum $\alpha$]}
For the compound secure groupcast problem (to any $N \in [1:K-1]$ of $K$ receivers), when key storage is minimized, $\alpha = 1$, broadcast bandwidth $\beta = \min(N, K-N+1)$ is achievable for any $N, K$ and is optimal when $N=2$ or $K-1$.
\begin{eqnarray}
&& \min\{ \beta~ |~ (\alpha=1, \beta) \in \mathcal{C} \} \leq \min(N, K-N+1).\\
&& \min\{ \beta~ |~ (\alpha=1, \beta) \in \mathcal{C} \} = 2, ~\mbox{when}~N=2, K-1 (K \geq 3).
\end{eqnarray}
\end{theorem}


The converse proof (when $\min(N, K-N+1) = 2$) follows from Theorem \ref{thm:ab2} (refer to the proof of Theorem \ref{thm:ab2} presented later in Section \ref{sec:ab2}). Note that $N \geq 2$ and $K \geq 3$.
\begin{eqnarray}
(\ref{eq:53}) \Rightarrow \alpha + \beta \geq 3 ~\overset{\alpha=1}{\Longrightarrow}~ \beta \geq 2.
\end{eqnarray} 


The achievability proof of Theorem \ref{thm:a1} is presented in Section \ref{sec:a1}. A proof outline is as follows.
The case where $\min(N, K-N+1) = N$ is trivial as the independent key solution will work (refer to Fig.~\ref{fig:mot}.2). We only need to consider the case where $\min(N, K-N+1) = K-N+1$. Here the key of each receiver is $1$ generic linear combination of $K-N+1$ uniform i.i.d. basis key symbols, i.e., we operate over a $K-N+1$ dimensional key space and each receiver is assigned a $1$ dimensional subspace. The transmit signal $X$ for any $N$ qualified receivers is a length $K-N+1$ vector, where each element is a sum of the message $W$ scaled by a constant and a basis key symbol. To ensure that the $K-N$ eavesdropping receivers learn nothing about the message, we precode the message to the direction that is orthogonal to the key space of each eavesdropping receiver (by choosing the $K-N+1$ constants before $W$ in $X$). This precoding vector  can be chosen as the null space of the $K-N$ dimensional subspace seen by all the eavesdropping receivers, which exists and has $1$ dimension over a $K-N+1$ dimensional space. Further, the subspaces are generic so that the projection of the precoded message to the $1$ dimensional subspace held by any qualified receiver is not empty, and correctness follows. It turns out that MDS matrices suffice for the above coding scheme. An example is given below to illustrate this idea.

\begin{example}\label{ex:a1}
Continuing from Fig.~\ref{fig:mot}.2, we consider the compound secure groupcast problem when $N=3, K=4$ and give a scheme that achieves $\alpha = 1, \beta = 2$. The generic key assignment is as follows. Define $s_1, s_2$ as $2$ uniform i.i.d. symbols from $\mathbb{F}_3$.
\begin{eqnarray}
Z_1 = s_1, ~Z_2 = s_2, ~Z_3 = s_1+s_2, ~Z_4 = s_1+2s_2.
\end{eqnarray}
Suppose receivers 1 to 3 are qualified. The transmit signal is set as
\begin{eqnarray}
X_{\{1,2,3\}} = \left[
\begin{array}{c}
-2\\
1
\end{array}
\right] W +
\left[ \begin{array}{c}
s_1 \\
s_2
\end{array}
\right]
\end{eqnarray}
where for $W$, the precoding vector $[-2, ~1]^T$ is orthogonal to the key space of the eavesdropping Receiver $4$, $[1, ~2]$, so that security is guaranteed. For correct decoding, take Receiver $3$ as an example, who will project the transmit signal to his key space.
\begin{eqnarray}
(-2W + s_1) + (W + s_2) = -W + s_1 + s_2 = - W + Z_3
\end{eqnarray}
so that the knowledge of $Z_3$ ensures that $W$ can be decoded with no error. The transmit signals for other cases are designed similarly.
\begin{eqnarray}
X_{\{1,2,4\}} = \left[
\begin{array}{c}
-1\\
1
\end{array}
\right] W +
\left[ \begin{array}{c}
s_1 \\
s_2
\end{array}
\right],~
X_{\{1,3,4\}} = \left[
\begin{array}{c}
1\\
0
\end{array}
\right] W +
\left[ \begin{array}{c}
s_1 \\
s_2
\end{array}
\right],~
X_{\{2,3,4\}} = \left[
\begin{array}{c}
0\\
1
\end{array}
\right] W +
\left[ \begin{array}{c}
s_1 \\
s_2
\end{array}
\right].
\end{eqnarray}
Finally, key storage $\alpha =1$ and broadcast bandwidth $\beta = 2 = K-N+1$ are achieved, as desired.
\end{example}



We next improve the result in Theorem \ref{thm:a1} for the simplest setting where the optimal broadcast bandwidth for the minimum key storage extreme point ($\alpha=1$) is open. For this setting ($N=3, K=5$), we have the following upper and lower bounds.

\begin{theorem}\label{thm:53}
{\normalfont[$N=3, K=5$]}
For the compound secure groupcast problem (to any $3$ of $5$ receivers), when key storage is minimized, $\alpha = 1$, the 
minimum broadcast bandwidth satisfies $2.5 \leq \beta \leq 2.9$.
\begin{eqnarray}
2.5 \leq \min\{ \beta~ |~ (\alpha=1, \beta) \in \mathcal{C} \} \leq 2.9, ~\mbox{when}~N=3, K=5.
\end{eqnarray}
\end{theorem}

The converse proof of Theorem \ref{thm:53} is deferred to Section \ref{sec:53}, 
where we translate the lower bound on broadcast bandwidth to a lower bound on joint key size for $3$ receivers, $H(Z_1, Z_2, Z_3) \geq 2.5 L_W$.
The achievable scheme is presented now, where we improve the achievability of $\beta = 3$ in Theorem \ref{thm:a1} to that of $\beta = 2.9$ by introducing some correlation into the generic spaces used in Theorem \ref{thm:a1}.

We first present a scalar linear scheme that achieves average broadcast bandwidth $2.9$. The keys are assigned as
\begin{eqnarray}
Z_1 = s_1, ~Z_2 = s_2, ~Z_3 = s_3, ~Z_4 = s_4, ~Z_5 = s_1 + s_2
\end{eqnarray}
where $s_1, s_2, s_3, s_4$ are $4$ uniform i.i.d. symbols from $\mathbb{F}_3$. There are $\binom{K}{N} = \binom{5}{3} = 10$ distinct choices of the qualified receivers, out of which $1$ can be achieved with $2$ symbols of broadcast signal and the remaining cases require $3$ symbols of broadcast signal, to groupcast $L_W = 1$ message symbol. The design principle is the same as that of the achievability proof of Theorem \ref{thm:a1}. Specifically, $\forall i \in \{3,4\}$, $\forall j \in \{1,2,5\}$,
\begin{eqnarray}
&& X_{\{1,2,i\}} = \left[
\begin{array}{c}
W + s_1\\
-W + s_2\\
W + s_i
\end{array}
\right],~
X_{\{1,2,5\}} = \left[
\begin{array}{c}
W + s_1\\
W + s_2
\end{array}
\right],~
X_{\{j,3, 4\}} = \left[
\begin{array}{c}
W + Z_j \\
 W+ s_3 \\
 W + s_4
 \end{array}
\right],\\
&& X_{\{1,i,5\}} = \left[
\begin{array}{c}
W + s_1\\
s_2\\
W + s_i\\
\end{array}
\right],~
X_{\{2,i, 5\}} = \left[
\begin{array}{c}
s_1\\
W + s_2\\
W + s_i
\end{array}
\right]
\end{eqnarray}
where correctness and security are easy to verify. 
Next, we symmetrize the above scheme by applying it to any all permutations of the $5$ receivers and concatenation, i.e., the size of each of $W, Z_k, X_{\mathcal{Q}}$ is scaled by $5!$ and now broadcast bandwidth becomes symmetric and $\beta = 0.9\times 3 + 0.1\times 2 = 2.9$. The achievability proof is complete now.

\subsection{Key Storage and Broadcast Bandwidth Region} \label{sec:ab2}
In this section, we present our results on the capacity region $(\alpha, \beta)$.

\begin{figure}[h]
\begin{center}
\includegraphics[width=3 in]{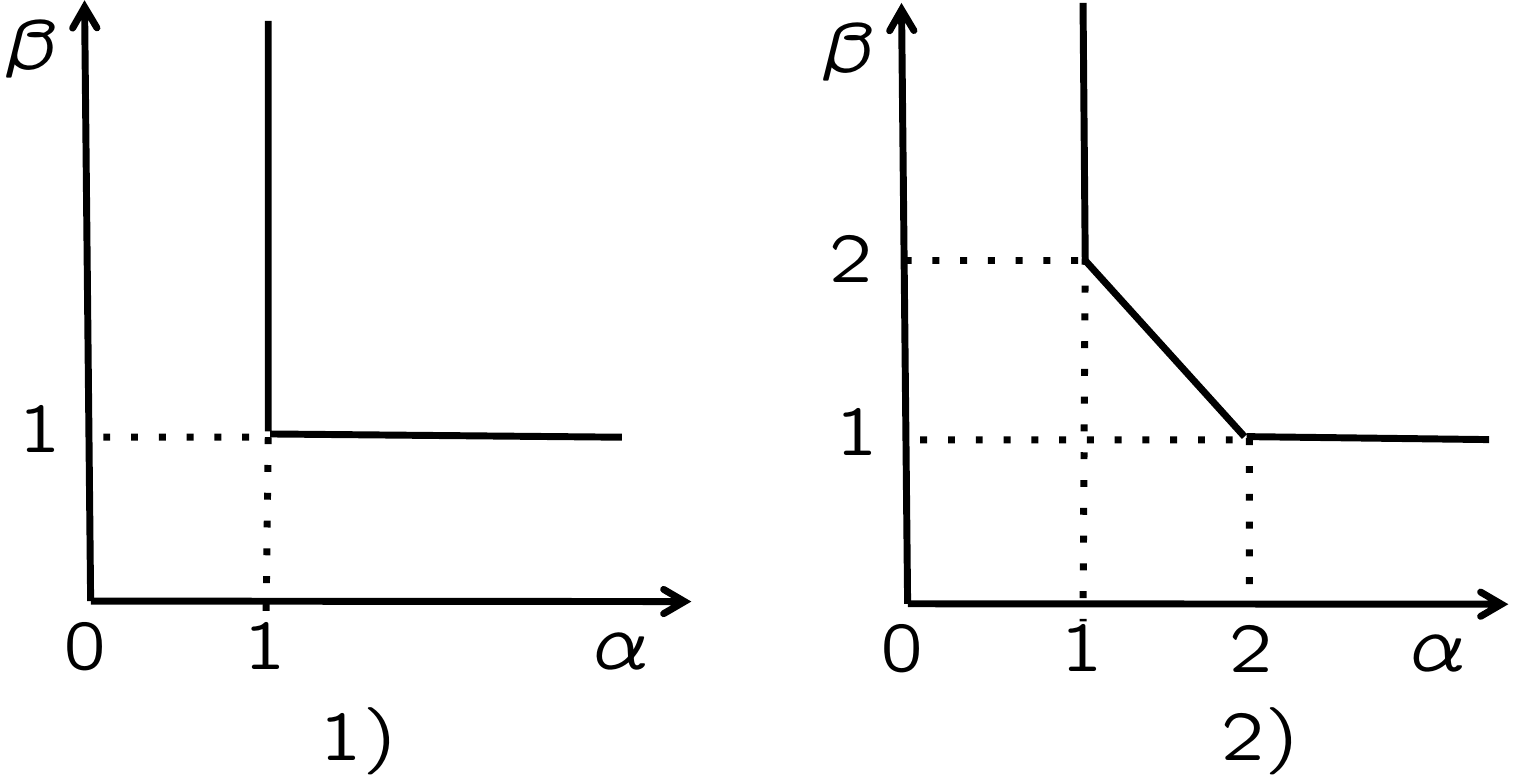}
\caption{\small The key storage and broadcast bandwidth region $(\alpha, \beta)$ of compound secure groupcast - 1) when $N=1$ and $K \geq 2$ is arbitrary, and 2) when $N=2$ and $K \geq 3$ is arbitrary.}
\label{fig:region}
\end{center}
\end{figure}

When $N=1$, the capacity region is characterized as $\mathcal{C} = \{(\alpha, \beta): \alpha \geq 1, \beta \geq 1\}$ and is plotted in Fig.~\ref{fig:region}.1. Here we only have one extreme point, where the minimum key storage and the minimum broadcast bandwidth are simultaneously attained. The proof is immediate, e.g., converse is proved in (\ref{eq:e1}), (\ref{eq:e2}) and achievability may follow from either Theorem \ref{thm:b1} or Theorem \ref{thm:a1}.

\vspace{0.05in}
\vspace{-0.2in}

Next we consider the case where $N=2$. It turns out that the capacity region is characterized fully by the two extreme points considered in previous sections (see Fig.~\ref{fig:region}.2). This result is stated in the following theorem.

\begin{theorem}\label{thm:ab2}
{\normalfont[$(\alpha, \beta)$ Region when $N=2$]}
For the compound secure groupcast problem (to any $N=2$ of $K$ receivers), the capacity region has and only has two extreme points - one corresponding to minimum key storage, $(\alpha, \beta) = (1,2)$ and the other corresponding to minimum broadcast bandwidth, $(\alpha, \beta) = (2,1)$.
\begin{eqnarray}
\mathcal{C} = \{(\alpha, \beta): \alpha + \beta \geq 3, \alpha \geq 1, \beta \geq 1\}, ~\mbox{when $N=2$ and $K \geq 3$ is arbitrary}. \label{eq:53}
\end{eqnarray}
\end{theorem}

{\it Proof:} Achievability of the two extreme points follows from Theorem \ref{thm:b1} and Theorem \ref{thm:a1}, respectively. For the converse, we only need to prove $\alpha + \beta \geq 3$ as $\alpha \geq 1$ has been proved in (\ref{eq:e1}) and $\beta \geq 1$ has been proved in (\ref{eq:e2}). The proof of $\alpha + \beta \geq 3$ is presented next, which is a simple consequence of the bounds in Theorem \ref{thm:con}. Consider (\ref{rate_con}) and set $q = 1, e = 2$.
\begin{eqnarray}
 L_W \leq H(Z_1 | Z_2). \label{eq:r1}
\end{eqnarray}
Consider (\ref{com_con}) and set $\mathcal{Q} = \{1,2\}$.
\begin{eqnarray}
L_X &\geq& 2L_W - (H(Z_1) + H(Z_2) - H(Z_1, Z_2)) \\
&=& 2L_W -  H(Z_1) + H(Z_1 | Z_2) \\
&\overset{(\ref{eq:r1})}{\geq}& 2L_W - L_Z + L_W \\
\Rightarrow \alpha + \beta = (L_Z + L_X)/L_W &\geq& 3. 
\end{eqnarray}
\hfill\QED

\section{Related Work}\label{sec:prior}
In this section, we discuss connections of compound secure groupcast to prior work in cryptography, mainly two lines - key predistribution and broadcast encryption. For related work on secure groupcast, we refer to \cite{Sun_SecureGroupcast}. 

The key (pre)-distribution problem refers to the assignment of a number of key variables (each belongs to a user) such that certain subsets (e.g., any $N$ of $K$ keys) can agree on an independent key that is known exclusively to the given subset of users. Key predistribution systems naturally provide an achievable scheme for compound secure groupcast, where the qualified receivers (users) can extract an exclusive secure key according to a key predistribution scheme and use this key to securely groupcast the message with one-time pad. Key predistribution based schemes turn out to be optimal for the key storage $\alpha$ and broadcast bandwidth $\beta$ tradeoff for compound secure groupcast, when broadcast bandwidth is minimized (refer to Theorem \ref{thm:b1}), and are sub-optimal in general, e.g., when key storage is minimized (refer to Theorem \ref{thm:a1}). Specifically, in establishing Theorem \ref{thm:b1}, we invoke elegant key predistribution schemes from early works in the field \cite{Blom1984, Matsumoto_Imai}.

Key predistribution has been extensively studied in cryptography since \cite{Blom1984, Matsumoto_Imai}, subject to both information theoretic security and computational security constriants. In particular, under information theoretic security, \cite{Blundo_Key} showed that for any $N$ out of $K$ users to agree on an independent key (i.e., the symmetric setting), the minimum storage per user is equal to $N$ times the key size. This result coincides with the minimum key storage result ($\alpha = N$) for the minimum broadcast bandwidth extreme point ($\beta = 1$) for compound secure groupcast. While key predistribution was originally studied in the symmetric any $N$ out of $K$ setting and in the non-interactive setting, interesting generalizations to allow asymmetric access structure and to allow communication among the users, so that storage size is reduced, have appeared in \cite{KPS_hierarchical} and \cite{Beimel_Chor}, respectively. For other key predistribution works under information theoretic security, we refer to survey papers \cite{Stinson_KPS, Blundo_KPS} and references therein. Computational security works can be found in cryptography textbooks \cite{Crypto_Handbook, Crypto_Stinson}.

Another line of related work is referred to as broadcast encryption \cite{Fiat_Naor_BE}, where the goal is to enable secure broadcasting to selected sets of receivers with the major focus on key management under dynamic changes of qualified users. Since its introduction \cite{Fiat_Naor_BE}, broadcast encryption has grown to a big umbrella that covers a wide array of themes, e.g., computational security (see e.g., \cite{attrapadung2003broadcast, boneh2005collusion}), key management with multicast trees (see e.g., \cite{Wong_Gouda_Lam, canetti1999multicast, Poovendran_Baras}), and user revocation (see e.g., \cite{naor2003protecting, kogan2006practical}).

The topic within broadcast encryption that is most related to our work is (information theoretically secure) one-time broadcast encryption schemes \cite{Blundo_BE, Stinson_KPS}, which is essentially the same as compound secure groupcast albeit it is not formulated in Shannon theoretic framework (with some minor difference). Entropy based approaches have been employed in  \cite{Blundo_BE, Desmedt_BE}, where the main focus is on the code construction and analyzing the storage requirement for key predistribution and one-time pad coding based schemes (which is efficient only near the $\beta = 1$ extreme point). The tradeoff between key storage and broadcast bandwidth has also been noted \cite{Stinson_KPS, Blundo_Tradeoff}, where various tools from coding theory, design theory, and secret sharing have been used to construct several classes of achievable schemes while the information theoretic optimality of the proposed schemes is largely unknown due to the lack of converse (notably one converse bound appeared in Theorem 3 of \cite{Blundo_Tradeoff} and it is equivalent to $\alpha + \beta \geq 3$ in our notation (proof omitted though \cite{Blundo_Tradeoff})).
In contrast, in this work we focus on establishing the optimality of extreme points and the capacity region of elemental systems, e.g., $N=2$. The schemes in \cite{Stinson_KPS, Blundo_Tradeoff} could be useful for future studies on the overall $(\alpha, \beta)$ tradeoff of larger compound secure groupcast systems (and generalizations to arbitrary access structure that goes beyond the symmetric any $N$ of $K$ setting and to include colluding eavesdropping receivers).


\section{Proofs}
\subsection{Proof of Theorem \ref{thm:con}}\label{sec:con}
First, consider (\ref{rate_con}). For any $q, e \in [1:K], q \neq e$, suppose Receiver $q$ is qualified and Receiver $e$ is not (i.e., eavesdropping), which is possible for any $q, e$ in the compound secure groupcast problem. Consider any qualified set of receivers, $\mathcal{Q}$ such that $q \in \mathcal{Q}, e \notin \mathcal{Q}, |\mathcal{Q}| = N$.
\begin{eqnarray}
L_W = H(W) &\overset{(\ref{wz_ind})}{=}& H(W | Z_e) \\
&\overset{(\ref{corr})}{=}& I(W; X_{\mathcal{Q}}, Z_q | Z_e) \\
&\overset{(\ref{sec})}{=}& I(W; Z_q | Z_e, X_{\mathcal{Q}}) \leq H(Z_q | Z_e).
\end{eqnarray}

Second, consider (\ref{com_con}). Without loss of generality, we set $\mathcal{Q} = \{1,\cdots, Q\}$, $Q \leq N$ and suppose receivers $1$ to $N$ are qualified.
\begin{eqnarray}
L_X \geq H(X_{[1:N]}) &\geq& I(X_{[1:N]}; W, Z_1, \cdots, Z_Q) \\
&=& I(X_{[1:N]}; W, Z_1) + \sum_{q=1}^{Q-1} I(X_{[1:N]}; Z_{q+1} | W, Z_1,\cdots, Z_q) \\
&\overset{(\ref{wz_ind})}{\geq}& I(X_{[1:N]}; W | Z_1) + \sum_{q=1}^{Q-1} I(X_{[1:N]}, W; Z_{q+1} | Z_1,\cdots, Z_q)\\
&\overset{(\ref{wz_ind})}{=}& I(X_{[1:N]}, Z_1; W) + \sum_{q=1}^{Q-1} \Big( I(X_{[1:N]}, W, Z_1,\cdots, Z_q; Z_{q+1}) \notag\\
&&-~ I(Z_1,\cdots, Z_q; Z_{q+1}) \Big)\\
&\overset{(\ref{corr})}{\geq}& H(W) + \sum_{q=1}^{Q-1} \Big(I(W; Z_{q+1}|X_{[1:N]}) - I(Z_1,\cdots, Z_q; Z_{q+1}) \Big) \\
&\overset{(\ref{corr})}{=}& H(W) + \sum_{q=1}^{Q-1} \Big(H(W|X_{[1:N]}) - I(Z_1,\cdots, Z_q; Z_{q+1}) \Big) \\
&\overset{(\ref{sec})}{=}& H(W) + \sum_{q=1}^{Q-1} \Big(H(W) - I(Z_1,\cdots, Z_q; Z_{q+1}) \Big) \\
&\overset{}{=}& QL_W - \Big(\sum_{q=1}^Q H(Z_q) - H(Z_1, \cdots, Z_Q) \Big).
\end{eqnarray}

\subsection{Proof of Theorem \ref{thm:b1}: $\alpha_{\min} = N$ when $\beta = 1$} \label{sec:b1}
We present the achievability proof and the converse proof in the following two sections.
\subsubsection{Achievability}
We show that $\alpha = N, \beta = 1$ is achievable. The coding scheme is as follows.

Denote by $s_1, \cdots, s_{N+1}$ a basis set of independent uniform symbols from $\mathbb{F}_p$, where $p$ is any prime power that is greater than or equal to $K$. Define ${\bf s} = [s_1, \cdots, s_{N+1}] \in \mathbb{F}_p^{1\times (N+1)}$. The keys are assigned according to
\begin{eqnarray}
&& Z_k = {\bf V}_k {\bf s}^T, ~\forall k \in [1:K] \\
&& \mbox{where}~ {\bf V}_k = \left[
\begin{array}{ccccccc}
1 & v_k & 0 & \cdots & 0 \\
0 & 1 & v_k & \cdots & 0 \\
\vdots & \cdots & \vdots & \cdots & 0 \\
0 & 0 & \cdots & 1 & v_k
\end{array}
\right]_{N \times (N+1)}\\
&& \mbox{and}~ v_1, \cdots, v_{K}~\mbox{are distinct elements from}~\mathbb{F}_p.
\end{eqnarray}
Note that the above elegant construction was introduced by Matsumoto and Imai \cite{Matsumoto_Imai}.

For any set of $N$ qualified receivers, $\mathcal{Q} = \{q_1, \cdots, q_N\} \subset [1:K]$, the transmit signal is 
\begin{eqnarray}
&& X_{\mathcal{Q}} = W + z_{\mathcal{Q}} = W + {\bf v}_{\mathcal{Q}} {\bf s}^T \\
&& \mbox{where}~{\bf v}_{\mathcal{Q}} = [1~~ \sum_{i} v_{q_i} ~~\sum_{i,j, i\neq j} v_{q_i} v_{q_j} ~~\cdots ~~\prod_{i}v_{q_i}]_{1 \times (N+1)}.
\end{eqnarray}
Note that the $n$-th element of ${\bf v}_{\mathcal{Q}}, n \in [1:N+1]$ has degree $n-1$ and is the sum of the product of all distinct $n-1$ terms from $v_{q_1}, \cdots, v_{q_N}$.
Note also that $W \in \mathbb{F}_p, X_{\mathcal{Q}} \in \mathbb{F}_p$.

We verify correctness. To this end, we show that every qualified receiver can recover ${\bf v}_{\mathcal{Q}}{\bf s}^T$ such that $W$ can be decoded with no error. For any $q \in \mathcal{Q}$,
\begin{eqnarray}
{\bf v}_{\mathcal{Q}} = {\bf V}_q(1, :) + {\bf V}_q(2, :) \sum_{i\neq q} v_{q_i} + {\bf V}_q(3, :) \sum_{i,j, i\neq j, i\neq q, j\neq q} v_{q_i} v_{q_j} + \cdots + {\bf V}_q(N, :) \prod_{i\neq q} v_{q_i}.
\end{eqnarray}
Hence, ${\bf v}_{\mathcal{Q}}$ is a linear combination of the rows of ${\bf V}_q$ and correctness holds. We consider security next. Consider any eavesdropping Receiver $e \in [1:K]\backslash\mathcal{Q}$.
\begin{eqnarray}
&& {\bf V}_{\mathcal{Q} \cup \{e\}} = \left[
\begin{array}{c}
{\bf V}_e \\
{\bf v}_{\mathcal{Q}} 
\end{array}
\right] = \left[
\begin{array}{ccccccc}
1 & v_e & 0 & \cdots & 0 \\
0 & 1 & v_e & \cdots & 0 \\
\vdots & \cdots & \vdots & \cdots & 0 \\
0 & 0 & \cdots & 1 & v_e \\
1 & \sum_{i} v_{q_i} & \sum_{i,j, i\neq j} v_{q_i} v_{q_j} & \cdots & \prod_{i}v_{q_i}
\end{array} 
\right]_{(N+1) \times (N+1)} \label{eq:qe}\\
&\Rightarrow& \det({\bf V}_{\mathcal{Q} \cup \{e\}}) = (-1)^N \prod_{i} (v_e - v_{q_i}) \neq 0 ~\mbox{as $v_k$ are distinct.}\label{eq:vqe}
\end{eqnarray}
Note that we have omitted the detailed steps of the derivation of the determinant formula (\ref{eq:vqe}), which is straightforward to verify. We show that the security constraint (\ref{sec}) is satisfied.
\begin{eqnarray}
I(W; X_{\mathcal{Q}}, Z_e) &=& H(X_{\mathcal{Q}}, Z_e) - H(X_{\mathcal{Q}}, Z_e | W) \\
&\overset{(\ref{wz_ind})(\ref{eq:qe})}{\leq}& (N+1) - H({\bf V}_{\mathcal{Q} \cup \{e\}} {\bf s}^T) \overset{(\ref{eq:vqe})}{=} 0. 
\end{eqnarray}
Finally, each key $Z_k$ has $N$ symbols and each broadcast signal $X_{\mathcal{Q}}$ has $1$ symbol, so we have achieved $\alpha = N, \beta = 1$. The achievability proof of Theorem \ref{thm:b1} is thus complete.

\subsubsection{Converse}
We prove that when $\beta = 1$, the inequality $\alpha \geq N$ holds.

Let us start with a useful lemma.
\begin{lemma}\label{lemma:b1}
When $\beta = 1$, for any $\mathcal{Q} \subset [1:K], |\mathcal{Q}| = N$, we have
\begin{eqnarray}
H(X_{\mathcal{Q}}) = H(X_{\mathcal{Q}} | W)&=& L_X = L_W, \label{eq:b1_xsize}\\
H(X_{\mathcal{Q}} | W, Z_q) &=& 0, \forall q \in \mathcal{Q}, \label{eq:b1_xdet}\\
I(X_{\mathcal{Q}} ; Z_e) = I(X_{\mathcal{Q}} ; Z_e | W) &=& 0, \forall e \notin \mathcal{Q}. \label{eq:b1_zind}
\end{eqnarray}
\end{lemma}

\begin{remark}
The interpretation of Lemma \ref{lemma:b1} is that when $\beta = 1$, every qualified set must agree on a key, and we essentially have to use one-time pad coding with such a key (see (\ref{eq:b1_xsize})(\ref{eq:b1_xdet})). Further, the key of any qualified set must be secure to any eavesdropping receiver (see (\ref{eq:b1_zind})).
\end{remark}

{\it Proof:} First consider (\ref{eq:b1_xsize}) and (\ref{eq:b1_xdet}). Consider any $q \in \mathcal{Q}$. Note that $\beta = L_X/L_W = 1$.
\begin{eqnarray}
L_X = L_W = H(W) &\overset{(\ref{wz_ind})}{=}& H(W | Z_q) \\
&\overset{(\ref{corr})}{=}& I(W; X_{\mathcal{Q}} | Z_q) \\
&\leq& H(X_{\mathcal{Q}}) - H(X_{\mathcal{Q}} | W, Z_q) \\
&\leq& H(X_{\mathcal{Q}}) \leq L_X \\
\Rightarrow ~~~~~~~~~~~~~ H(X_{\mathcal{Q}})  &=& L_X = L_W \\
	H(X_{\mathcal{Q}} | W, Z_q) &=& 0\\ 
\Rightarrow~~~~~~~~~ H(X_{\mathcal{Q}} | W) &\overset{(\ref{sec})}{=}& H(X_{\mathcal{Q}}) = L_X = L_W.
\end{eqnarray}


Then consider (\ref{eq:b1_zind}). Consider any $e \notin \mathcal{Q}$ and any $q \in \mathcal{Q}$. Note that $\beta = L_X/L_W = 1$.
\begin{eqnarray}
H(X_{\mathcal{Q}} | Z_e) &\geq& H(X_{\mathcal{Q}} | Z_e, Z_q) \\
&=& I(X_{\mathcal{Q}}; W | Z_e, Z_q) + H(X_{\mathcal{Q}} | W, Z_e, Z_q)\\
&\overset{(\ref{corr})(\ref{eq:b1_xdet})}{=}& H(W | Z_e, Z_q) \\
&\overset{(\ref{wz_ind})}{=}& H(W) \\
&\overset{(\ref{h1})}{=}& L_W \\
&=& L_X \\
&\geq& H(X_{\mathcal{Q}})\\
\Rightarrow~ I(X_{\mathcal{Q}} ; Z_e) &=& 0 ~~\mbox{as mutual information is non-negative} \label{eq:b1_xt}\\
 I(X_{\mathcal{Q}} ; Z_e | W) &\overset{(\ref{sec})}{=}& I(X_{\mathcal{Q}}; Z_e, W) \\
 &\overset{(\ref{eq:b1_xt})}{=}& I(X_{\mathcal{Q}}; W | Z_e)\\
 &\overset{(\ref{sec})}{=}& 0.
\end{eqnarray}
\hfill\QED

Next, we consider only the first $N+1$ receivers. Note that $K \geq N+1$ and removing users cannot enlarge the capacity region. Consider all qualified sets that include Receiver $N+1$ and there are $N$ such qualified sets, i.e., $\mathcal{Q}_1 = [1:N+1]\backslash \{1\}, \cdots, \mathcal{Q}_{N} = [1:N+1]\backslash \{N\}$.
\begin{eqnarray}
L_Z \geq H(Z_{N+1}) &\overset{(\ref{wz_ind})}{=}& H(Z_{N+1} | W) \\
&\geq& I(Z_{N+1}; X_{\mathcal{Q}_1}, X_{\mathcal{Q}_2}, \cdots, X_{\mathcal{Q}_N} | W)\\
&=& \sum_{i=1}^N I(Z_{N+1}; X_{\mathcal{Q}_i} | W, X_{\mathcal{Q}_1},\cdots, X_{\mathcal{Q}_{i-1}})  \\
&\overset{(\ref{eq:b1_xdet})}{=}&  \sum_{i=1}^N H(X_{\mathcal{Q}_i} | W, X_{\mathcal{Q}_1},\cdots, X_{\mathcal{Q}_{i-1}}) \label{eq:b11}\\
&=&  \sum_{i=1}^N \Big( H(X_{\mathcal{Q}_i} | W) - I(X_{\mathcal{Q}_i}; X_{\mathcal{Q}_1},\cdots, X_{\mathcal{Q}_{i-1}} | W) \Big)\\
&\geq&  \sum_{i=1}^N \Big( H(X_{\mathcal{Q}_i} | W) - I(X_{\mathcal{Q}_i}; Z_i, X_{\mathcal{Q}_1},\cdots, X_{\mathcal{Q}_{i-1}} | W) \Big)\\
&\overset{ (\ref{eq:b1_xdet})}{=}& \sum_{i=1}^N H(X_{\mathcal{Q}_i} | W) - \sum_{i=1}^N I(X_{\mathcal{Q}_i}; Z_i | W) \label{eq:b12}\\
&\overset{(\ref{eq:b1_xsize}) (\ref{eq:b1_zind})}{=}& N L_W - 0 \label{eq:b13}\\
\Rightarrow~ \alpha = L_Z/L_W &\geq& N
\end{eqnarray}
where (\ref{eq:b11}) follows from the observation that $N+1 \in \mathcal{Q}_i$. In (\ref{eq:b12}), we use the fact that $i \in \mathcal{Q}_1,\cdots,\mathcal{Q}_{i-1}, i \geq 2$, and in (\ref{eq:b13}), we use the fact that $i \notin \mathcal{Q}_i$.
The converse proof of Theorem \ref{thm:b1} is now complete.

\subsection{Achievability Proof of Theorem \ref{thm:a1}: $\beta \leq \min(N, K-N+1)$ when $\alpha = 1$} \label{sec:a1}
We present a coding scheme that achieves $\alpha = 1, \beta = \min(N, K-N+1)$. We have two cases, depending on $\min(N, K-N+1)$ is equal to $N$ or $K-N+1$.

First, consider the case where $\min(N, K-N+1) = K - N + 1$, and we show that $\beta = K-N+1$ is achievable. 
Define ${\bf s} = [s_1, \cdots, s_{K-N+1}] \in \mathbb{F}_p^{1\times (K-N+1)}$, where $s_1, \cdots, s_{K-N+1}$ are $K-N+1$ uniform i.i.d. symbols from $\mathbb{F}_p$ and $p$ is a prime power that is greater than or equal to $K$. The keys are assigned as
\begin{eqnarray}
Z_k = {\bf V}(k,:) {\bf s}^T
\end{eqnarray}
where ${\bf V}(k,:) \in \mathbb{F}_p^{1 \times (K+N-1)}$ is the $k$-th row of the matrix ${\bf V}$, and
\begin{eqnarray}
\mbox{${\bf V}$ is a $K\times (K-N+1)$ MDS matrix}.
\end{eqnarray}
For our scheme, the only requirement on the field size $p$ is that an MDS matrix ${\bf V}$ exists over $\mathbb{F}_p$. So $p \geq K$ suffices as we may set ${\bf V}$ as the Vandermonde matrix.
For any $N$ receivers from the qualified set $\mathcal{Q} \subset [1:K], |\mathcal{Q}| = N$, the transmit signal is set as
\begin{eqnarray}
X_{\mathcal{Q}} = {\bf v}_W^T W + {\bf s}^T
\end{eqnarray} 
where $W \in \mathbb{F}_p$ is $L_W = 1$ symbol, $X_{\mathcal{Q}}\in \mathbb{F}_p^{(K-N+1) \times 1}$ is a column vector and ${\bf v}_W^T \in \mathbb{F}_p^{(K-N+1)\times 1}$ is chosen such that
\begin{eqnarray}
{\bf V}([1:K]\backslash\mathcal{Q}, :) \times {\bf v}_W^T = \vec{0}_{(K-N) \times 1}, \label{eq:b1az}
\end{eqnarray} 
i.e., ${\bf v}_W^T$ may be set as the right null space of the $(K-N) \times (K-N+1)$ matrix ${\bf V}([1:K]\backslash\mathcal{Q}, :)$. Such a column vector ${\bf v}_W^T$ exists because ${\bf V}$ is MDS and ${\bf V}([1:K]\backslash\mathcal{Q}, :)$ has rank $K-N$. Further,
\begin{eqnarray}
{\bf V}(q, :) \times {\bf v}_W^T  \neq 0, \forall q \in \mathcal{Q} \label{eq:b1a}
\end{eqnarray}
because otherwise ${\bf V}([1:K]\backslash\mathcal{Q} \cup \{q\}, :) \times {\bf v}_W^T= 0$, which leads to that a $(K-N+1) \times (K-N+1)$ sub-matrix of ${\bf V}$, ${\bf V}([1:K]\backslash\mathcal{Q} \cup \{q\}, :) $, is rank deficient, violating the fact that ${\bf V}$ has been set as an MDS matrix.

To show that zero error decoding is guaranteed, consider any qualified Receiver $q \in \mathcal{Q}$, who will project the received signal $X_{\mathcal{Q}}$ to his key space.
\begin{eqnarray}
{\bf V}(q, :) \times X_{\mathcal{Q}} = {\bf V}(q, :) \times {\bf v}_W^T W+ {\bf V}(q, :) \times {\bf s}^T
= 
\Big( {\bf V}(q, :) \times {\bf v}_W^T \Big) W + Z_q
\end{eqnarray}
such that from (\ref{eq:b1a}), ${\bf V}(q, :) \times {\bf v}_W^T$ is a non-zero scalar and $W$ can be recovered. We now verify security. Consider any eavesdropping Receiver $e \in [1:K] \backslash \mathcal{Q}$.
\begin{eqnarray}
I(W; X_{\mathcal{Q}}, Z_e) &=& I(W; {\bf v}_W^T W + {\bf s}^T, {\bf V}(e,:) {\bf s}^T) \\
&\overset{(\ref{eq:b1az})}{=}& I(W; {\bf v}_W^T W + {\bf s}^T) \label{eq:b1ax} \\
&=& H({\bf v}_W^T W + {\bf s}^T) - H({\bf v}_W^T W + {\bf s}^T | W) \\
&\overset{(\ref{wz_ind})}{\leq}& (K + N - 1) - H({\bf s}^T) = 0
\end{eqnarray}
where $(\ref{eq:b1ax})$ follows from the fact that ${\bf V}(e,:)  {\bf s}^T$ is a deterministic function of ${\bf v}_W^T W + {\bf s}^T$, i.e.,
\begin{eqnarray}
{\bf V}(e,:) \times \Big( {\bf v}_W^T W + {\bf s}^T \Big) = \Big({\bf V}(e,:) \times {\bf v}_W^T \Big) W + {\bf V}(e,:) {\bf s}^T \overset{(\ref{eq:b1az})}{=} {\bf V}(e,:) {\bf s}^T.
\end{eqnarray}
Hence the security constraint is satisfied. Finally, we have achieved $\alpha = 1$ as each $Z_k$ has $1$ symbol and $\beta = K-N+1$ as each $X_{\mathcal{Q}}$ has $K+N-1$ symbols.

Second, consider the case where $\min(N, K-N+1) = N$. While the simple scheme that uses fully independent keys will work (refer to Fig.~\ref{fig:mot}.1), we present a scheme with smaller joint key size.

Define ${\bf s} = [s_1, \cdots, s_{N+1}] \in \mathbb{F}_p^{1 \times (N+1)}$, where $s_1, \cdots, s_{N+1}$ are $N+1$ uniform i.i.d. symbols from $\mathbb{F}_p$ and $p$ is a prime power that is greater than or equal to $K$. The keys are assigned as
\begin{eqnarray}
Z_k = {\bf V}(k,:) {\bf s}^T, ~\mbox{where \mbox{${\bf V}$ is a $K\times (N+1)$ MDS matrix}}
\end{eqnarray}
and the transmit signal $X_{\mathcal{Q}}, \forall \mathcal{Q} \subset [1:K], |\mathcal{Q}| = N$ is set as
\begin{eqnarray}
X_{\mathcal{Q}} = \vec{1}_{N\times 1} W + {\bf V}(\mathcal{Q},:) {\bf s}^T
\end{eqnarray}
where $\vec{1}$ is an $N\times 1$ column vector such that every element is $1$ and $X_{\mathcal{Q}}$ has dimension $N \times 1$. To see correctness, note that any qualified receiver $q \in \mathcal{Q}$ can decode $W$ with no error from $W + {\bf V}(q,:){\bf s}^T = W + Z_q$, which is one row of $X_{\mathcal{Q}}$. To see security, note that any eavesdropping receiver cannot learn anything about $W$ because
\begin{eqnarray}
I(W; X_{\mathcal{Q}}, Z_e) &=& H(X_{\mathcal{Q}}, Z_e) - H(X_{\mathcal{Q}}, Z_e | W) \\
&\overset{(\ref{wz_ind})}{\leq}& (N+1) - H({\bf V}(\mathcal{Q} \cup \{e\},:) {\bf s}^T) = 0
\end{eqnarray}
where the last step follows from the fact that ${\bf V}$ is an MDS matrix such that any $N+1$ rows have full rank. Finally, we have achieved $\alpha = 1$ and $\beta = N$, as desired.

The achievability proof of Theorem \ref{thm:a1} is now complete.

\subsection{Converse Proof of Theorem \ref{thm:53}: $\beta \geq 2.5$ when $\alpha = 1, N=3, K=5$}\label{sec:53}
Let us start with a useful lemma.
\begin{lemma}\label{lemma:53}
When $\alpha = 1$, for any $q, e \in [1:K], q \neq e$, we have
\begin{eqnarray}
H(Z_q) &=& L_W, \label{eq:53_z}\\
H(Z_q, Z_e) &=& 2 L_W. \label{eq:53_2z}
\end{eqnarray}
\end{lemma}

\begin{remark}
The interpretation of Lemma \ref{lemma:53} is that when $\alpha = 1$, the key at any receiver is uniform and the keys at any two receivers are independent.
\end{remark}

{\it Proof:} Note that $\alpha = L_Z/ L_W = 1$. From (\ref{rate_con}) in Theorem \ref{thm:con}, we have
\begin{eqnarray}
&& L_Z = L_W \leq H(Z_q | Z_e) \leq H(Z_q) \leq L_Z \\
&\Rightarrow& H(Z_q) = L_W ~\mbox{and symmetrically,}~ H(Z_e) = L_W \label{eq:53t}\\
&& H(Z_q | Z_e) = H(Z_q) \Rightarrow I(Z_q; Z_e) = 0 \Rightarrow H(Z_q, Z_e) = H(Z_q) + H(Z_e) \overset{(\ref{eq:53t})}{=} 2L_W.
\end{eqnarray}
\hfill\QED

We next show that $H(Z_1, Z_2, Z_3) \geq 2.5L_W$. To this end, we consider only the first $4$ receivers (which cannot help for the converse) and assume that Receiver $1$ and Receiver $3$ are qualified while Receiver $2$ and Receiver $4$ are eavesdropping. From submodularity of entropy functions, we have
\begin{eqnarray}
H(Z_1, Z_2, Z_3) + H(Z_1, Z_3, X_{\{1,3\}}, W) &\geq& H(Z_1, Z_3) + H(Z_1, Z_2, Z_3, X_{\{1,3\}}, W) \\
&\geq& H(Z_1, Z_3) + H(Z_2, X_{\{1,3\}}, W)\\
&\overset{(\ref{sec})}{=}& H(Z_1, Z_3) + H(Z_2, X_{\{1,3\}}) + H(W) \label{eq:53t1}\\
\mbox{Similarly,}~~H(Z_1, Z_2, Z_4) + H(Z_2, Z_4, X_{\{1,3\}}) &\geq& H(Z_2, Z_4) + H(Z_1, Z_2, Z_4, X_{\{1,3\}}) \\
&\overset{(\ref{corr})}{=}& H(Z_2, Z_4) + H(Z_1, Z_2, Z_4, X_{\{1,3\}}, W) \\
&\geq& H(Z_2, Z_4) + H(Z_4, X_{\{1,3\}}, W)\\
&\overset{(\ref{sec})}{=}& H(Z_2, Z_4) + H(Z_4, X_{\{1,3\}}) + H(W). \label{eq:53t2}
\end{eqnarray}
Adding (\ref{eq:53t1}) and (\ref{eq:53t2}), we have
\begin{eqnarray}
&& H(Z_1, Z_2, Z_3) + H(Z_1, Z_3, X_{\{1,3\}}, W) + H(Z_1, Z_2, Z_4) + H(Z_2, Z_4, X_{\{1,3\}}) \notag\\
&\geq& H(Z_1, Z_3) + H(Z_2, Z_4) + 2H(W) + H(Z_2, X_{\{1,3\}}) + H(Z_4, X_{\{1,3\}})\\
&\geq& H(Z_1, Z_3) + H(Z_2, Z_4) + 2H(W) + H(X_{\{1,3\}}) +  H(Z_2, Z_4, X_{\{1,3\}}) \\
&\overset{(\ref{h1}) (\ref{eq:53_2z})}{=}& 6 L_W + H(X_{\{1,3\}}) +  H(Z_2, Z_4, X_{\{1,3\}}).
\end{eqnarray}
Note that by symmetry, we may assume $H(Z_1, Z_2, Z_3) = H(Z_1, Z_2, Z_4)$ without loss of generality. Plugging this above, we have
\begin{eqnarray}
H(Z_1, Z_3, X_{\{1,3\}}, W) + 2H(Z_1,Z_2,Z_3) &\geq& 6 L_W + H(X_{\{1,3\}}). \label{eq:53t3}
\end{eqnarray}
A upper bound of $H(Z_1, Z_3, X_{\{1,3\}}, W)$ can be obtained as follows.
\begin{eqnarray}
2L_W + 2 H(X_{\{1,3\}}) &\overset{(\ref{eq:53_z})}{=}& H(Z_1) + H(X_{\{1,3\}}) + H(Z_3) + H(X_{\{1,3\}})\\
 &\geq& H(Z_1, X_{\{1,3\}}) + H(Z_3, X_{\{1,3\}}) \\
&\overset{(\ref{corr})}{=}& H(Z_1, X_{\{1,3\}}, W) + H(Z_3, X_{\{1,3\}}, W) \\
&\geq& H(X_{\{1,3\}}, W) + H(Z_1, Z_3, X_{\{1,3\}}, W) \\
&\overset{(\ref{sec})}{=}& H(X_{\{1,3\}}) + H(W) + H(Z_1, Z_3, X_{\{1,3\}}, W) \label{eq:53t4} \\
(\ref{eq:53t3}) + (\ref{eq:53t4}) \Rightarrow~ H(Z_1, Z_2, Z_3) &\geq& 2.5 L_W. \label{eq:53t5}
\end{eqnarray}
Finally, we invoke (\ref{com_con}) in Theorem \ref{thm:con} to translate the joint key size bound above to the desired broadcast bandwidth bound. Consider the $N=3, K=5$ compound secure groupcast problem and set $\mathcal{Q} = \{1,2,3\}$. From (\ref{com_con}), we have
\begin{eqnarray} 
L_X &\geq& 3 L_W - \Big(H(Z_1) + H(Z_2) + H(Z_3) - H(Z_1, Z_2, Z_3)\Big) \\
&\overset{(\ref{eq:53_z})}{=}& H(Z_1, Z_2, Z_3) \\
&\overset{(\ref{eq:53t5})}{\geq}& 2.5 L_W \\
\Rightarrow~ \beta = L_X/L_W &\geq& 2.5.
\end{eqnarray}

\subsubsection{Minimum Joint Key Size is 2.5 when $(N,K,\alpha,\beta) = (2,4,1,2)$}
In this section, we show that for the compound secure groupcast problem to any $N=2$ of $K=4$ receivers, when $(\alpha,\beta) =(1,2)$ (i.e., the minimum key storage extreme point), the minimum normalized joint key size $H(Z_1, Z_2, Z_3, Z_4)/L_W$ is $2.5$. We present this result because the minimum joint key size could be a useful auxiliary parameter, e.g., it is used to prove the broadcast bandwidth converse of Theorem \ref{thm:53}. In addition, the minimum joint key size is an interesting parameter by itself as it captures the minimum randomness resource required for compound secure groupcast. 

The converse $H(Z_1, Z_2, Z_3, Z_4) \geq H(Z_1, Z_2, Z_3) \geq 2.5 L_W$ follows from (\ref{eq:53t5}), proved in the previous section, and the achievability is presented below. The keys are assigned as
\begin{eqnarray}
Z_1 = (s_1, s_2), ~ Z_2 = (s_3, s_4), ~ Z_3 = (s_5, s_1+s_3), ~Z_4 = (s_2+s_4, s_1+s_2+s_5)
\end{eqnarray}
where $s_1, s_2, s_3, s_4, s_5$ are $5$ uniform i.i.d. symbols from any field $\mathbb{F}_p$. The message $W$ has $2$ symbols, $W = (W_1, W_2)$. The transmit broadcast signals are 
\begin{eqnarray}
X_{\{1,2\}} = \left[
\begin{array}{c}
W_1 + s_1\\
W_2 + s_2\\ \hline
-W_1 + s_3\\
-W_2 + s_4
\end{array}
\right],~
X_{\{1,3\}} = \left[
\begin{array}{c}
W_1 + s_1\\
W_2 + s_2\\ \hline
-W_1-W_2 + s_5\\
W_1 + s_1+s_3
\end{array}
\right],~
X_{\{1,4\}} = \left[
\begin{array}{c}
W_1 + s_1\\
W_2 + s_2\\ \hline
W_2 + s_2 + s_4\\
W_1 + W_2 + s_1+ s_2 + s_5
\end{array}
\right] \notag \\
X_{\{2,3\}} = \left[
\begin{array}{c}
W_1 + s_3\\
W_2 + s_4\\ \hline
W_2 + s_5\\
W_1 + s_1+ s_3
\end{array}
\right],~
X_{\{2,4\}} = \left[
\begin{array}{c}
W_1 + s_3\\
W_2 + s_4\\ \hline
W_2 + s_2 + s_4\\
-W_1 + s_1+ s_2 + s_5
\end{array}
\right],~
X_{\{3,4\}} = \left[
\begin{array}{c}
W_1 + s_5\\
W_2 + s_1 + s_3 \\ \hline
-W_2 + s_2 + s_4 \\
W_1 + s_1 + s_2 + s_5
\end{array}
\right]~
\end{eqnarray}
where the first (second) qualified receiver will use the first (last) two rows of $X$ to decode $W$ and security is guaranteed because the projection of $W$ in $X$ to the key space known by any eavesdropping receiver is empty. We finally calculate the performance of this scheme.
Note that $L_W = 2, L_Z = 2, L_X = 4, H(Z_1, Z_2, Z_3, Z_4, Z_5) = H(s_1, s_2, s_3, s_4, s_5) = 5$, so we have achieved 
\begin{eqnarray}
\alpha = L_Z/L_W = 1, ~\beta = L_X/L_W = 2, ~H(Z_1,Z_2,Z_3,Z_4,Z_5)/L_W = 2.5
\end{eqnarray}
and the proof is thus complete.

\section{Conclusion}
Motivated by the need to enable secure groupcast with demand uncertainty and inspired by related work in cryptography (especially broadcast encryption), we consider the compound secure groupcast problem that studies how to assign keys to efficiently and securely communicate with any $N$ of $K$ receivers through noiseless broadcasting, and focus on the tradeoff between key storage $\alpha$ and broadcast bandwidth $\beta$ from an information theoretic perspective.

Complete answers are found when broadcast bandwidth is minimized, i.e., when $\beta = 1$, the minimum key storage is $\alpha = N$, while the results are not tight when key storage is minimized, e.g., when $\alpha = 1$, broadcast bandwidth $\beta = \min(N, K-N+1)$ is achievable yet not optimal in general (i.e., settings with $N \geq 3, K \geq N+2$ are open).  Regarding the general $\alpha, \beta$ tradeoff, i.e., the $(\alpha, \beta)$ region, while $N=2$ cases are settled fully by the two extreme points where either $\alpha = 1$ or $\beta = 1$, settings with $N \geq 3$ remain open.

The solutions of this work mainly rely on generic spaces (matrices) and are found with an alignment (signal space) view of the problem (which also appears useful in several security and privacy primitives \cite{Li_Sun_CDS, Zhou_Sun_Fu, Sun_Jafar_PIR}). To further improve the achievable schemes, more structured spaces are in demand and remain an interesting future research direction.

\let\url\nolinkurl
\bibliographystyle{IEEEtran}
\bibliography{Thesis}
\end{document}